\documentclass{article}
\usepackage{spconf,amsmath,epsfig}
\usepackage{algorithm,amsfonts}
\usepackage{subfigure}
\usepackage{graphicx}
\usepackage{tabularx}
\usepackage{multirow}
\usepackage{array}

\title{Hyperspectral Image Denoising Based On Multi-Stream \\ Denoising Network}

\name{Yan Gao$^{1,2}$, Feng Gao$^{1,2,*}$, Junyu Dong$^{1,2}$ \thanks{This work was supported in part by the National Key Research and Development Program of China under Grant 2018AAA0100602, in part by the National Natural Science Foundation of China under Grant U1706218, and in part by the Key Research and Development Program of Shandong Province under Grant 2019GHY112048. (Email: gaofeng@ouc.edu.cn)}}
\address{$^1$College of Information Science and Engineering, Ocean University of China \\
$^2$ Institute of Marine Development, Ocean University of China }

\begin{document}

\maketitle

\begin{abstract}
Hyperspectral images (HSIs) have been widely applied in many fields, such as military, agriculture, and environment monitoring. Nevertheless, HSIs commonly suffer from various types of noise during acquisition. Therefore, denoising is critical for HSI analysis and applications. In this paper, we propose a novel blind denoising method for HSIs based on Multi-Stream Denoising Network (MSDNet). Our network consists of the noise estimation subnetwork and denoising subnetwork. In the noise estimation subnetwork, a multiscale fusion module is designed to capture the noise from different scales. Then, the denoising subnetwork is utilized to obtain the final denoising image. The proposed MSDNet can obtain robust noise level estimation, which is capable of improving the performance of HSI denoising. Extensive experiments on HSI dataset demonstrate that the proposed method outperforms four closely related methods.
\end{abstract}


\section{Introduction}

Hyperspectral images (HSIs) are typically composed of many spectral channels ranging from visible spectrum to infrared spectrum. HSIs can simultaneously acquire both spatial and spectral information, which provide richer scene information than other data sources. Nevertheless, HSIs are often affected by various types of noise because of imaging equipment and external environment in the process of acquisition, conversion, transmission, compression, and storage. Noise not only affects the visualization of HSIs, but also limits the subsequent analysis and processing of HSIs. Therefore, it is critical to remove the noise in HSIs before HSI analysis and processing.

To remove the noise interference, researchers have proposed many methods for HSIs denoising \cite{6165657} \cite{6819824}. Band -by-band denoising strategies are usually followed in traditional HSI denoising methods. Each band is considered as one 2-D image, such as block matching and 3-D filtering (BM3D) \cite{4271520} and weighted nuclear norm minimization (WNNM) \cite{Gu_2014_CVPR}. However, these strategies generally lead to large spectral distortions due to disregarding the spectral information. Different from these methods, block matching and 4-D filtering (BM4D) \cite{6253256} algorithm is 3-D image denoising method suitable for HSIs. However, it fails to take into account the inconsistency of the noise distribution between different bands. Therefore, BM4D does not perform well in the spectral domain.

\begin{figure*}[htb]
\begin{center}
\includegraphics [width=6in]{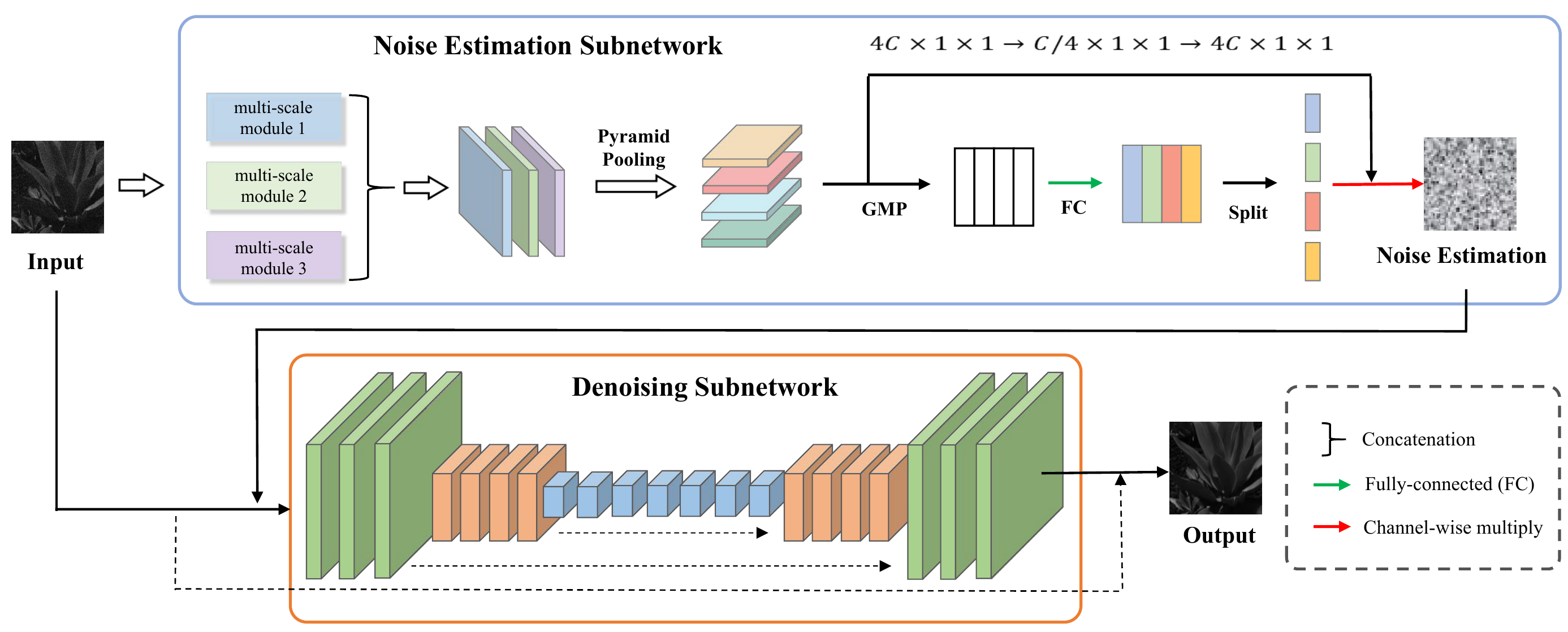}
\caption{Framework of the proposed image denoising method}
\end{center}
\end{figure*}

Recently, deep learning has achieved great success in image denoising due to its powerful capabilities of feature learning and nonlinear mapping. Image denoising models based on convolution neural networks (CNNs) develop rapidly. However, most of these methods remove the noise based on a specific noise level, and it is difficult to achieve the promising performance once the noise level changes. We often obtains the over-denoising or under-denoising results. To address the problem, blind denoising techniques are proposed. On the one hand, the denoising model is optimized by a very large training dataset which containing noisy images with various noise levels. However, it may be very tough for the network to learn all noise types at the same time. On the other hand, the noise estimation or noised label is introduced to guide the denoising process. Zhang et al. \cite{8365806} proposed a fast and flexible denoising network (FFDNet), which exhibited a relevant performance improvement in image denoising. FFDNet removes the complex noise by combining noisy image estimation and noise level estimation. However, unspecified noise level still deteriorates the performance. Therefore, it is challenging to design a generalized denoising model.

In this paper, we propose a blind denoising method for HSIs based on Multi-Stream Denoising Network (MSDNet), which can estimate the noise level autonomously instead of taking noise level as input. In MSDNet, noise estimation subnetwork is well-designed to produce the noise estimation, and then the denoising subnetwork is introduced to generate the final result. In particular, a multi-scale fusion module is developed to capture the noise at different scales in the noise estimation subnetwork. Experiments conducted on the HSIs dataset demonstrate that the proposed method is superior to other closely related methods.

\section{Methodology}

The framework of the proposed method is illustrated in Fig.1. It consists of two subnetworks: 1) noise estimation subnetwork; 2) denoising subnetwork. In the remainder of this section, more details about each subnetwork and the loss function will be described.

\subsection{Noise Estimation Subnetwork}

\begin{figure}[h]
\begin{center}
\includegraphics [width=3in]{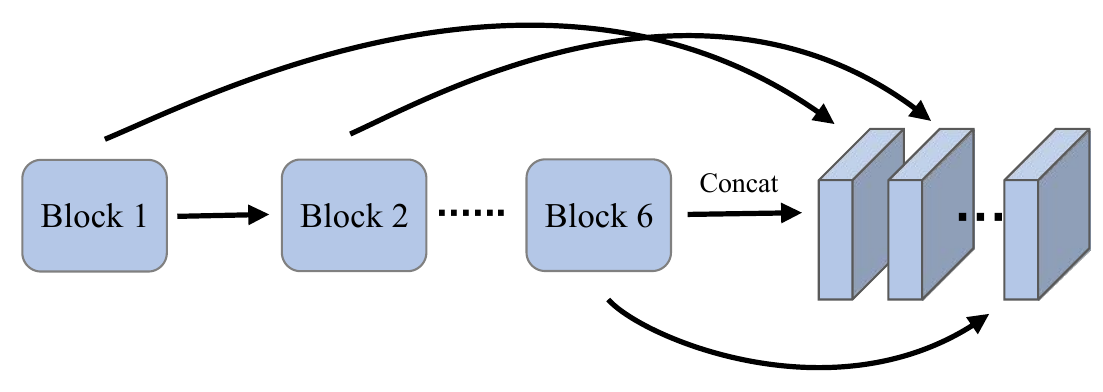}
\caption{Architecture of the multiscale module}
\end{center}
\end{figure}

To capture noise features, three multiscale modules with different kernel sizes are employed. Fig.2 shows the architecture of multiscale module. Each one consists of six blocks, and the output of each module can be defined as follows:
\begin{equation}
\textbf M_i=cat\lbrack B_1,B_2,\dots,B_6\rbrack,
\end{equation}
where $M_i,i=1,2,3$ denotes the $i$th multi-scale module, and $B_j,j=1,2,\dots,6$ denotes the $j$th block, and $cat(\cdot)$ is the concatenate operator in the channel dimension. 

As shown in Fig.1, the kernel sizes of the three multiscale modules are $3 \times 3$, $5\times5$, and $7 \times 7$, respectively. The outputs of three multiscale modules are concatenated together for the next step. We use the pyramid pooling module developed in \cite{Zhao_2017_CVPR} to further obtain the multiscale description. In order to fuse multiscale features for noise estimation, we concatenate four feature maps on the channel dimension to generate a feature map $P$ :
\begin{equation}
\textbf P=cat\lbrack f_1,f_2,f_3,f_4\rbrack,
\end{equation}
where $f_1,f_2,f_3,f_4$ denote four feature maps.

Afterwards, aiming to learn the relationship between the channels of feature maps, the attention mechanism is employed: Firstly, squeeze $P$ in the spatial domain to generate a vector $V\in \mathbb{R}^{4C \times 1 \times 1}$ by global max pooling. Then, $V$ is fed into two fully-connected layers to generate the weight vector $S$. Ultimately, channel-wise multiply is employed between the feature map $P$ and weight vector $S$:
\begin{equation}
\textbf n_i=f(P_i,S_i)\quad i=1,2,\dots,4C,
\end{equation}
where $n_i$ denotes the $i$th channel of noise level estimation $N$, and $f(\cdot)$ refers to channel-wise multiply between feature $P_i\in\mathbb{R}^{H\times W}$ and weight vector $S_i$ .

\subsection{Denoising Subnetwork}

In the denoising subnetwork, we employ a 16-layer UNet framework which takes both noise level estimation and noisy image as input to get the denoised image. All filter size of the network is $3 \times 3$ and the convolution layers are activated by ReLU function except the last one. The network obtains the denoised image $D$ by learning the residual mapping of the noisy image $Y$ as follows:
\begin{equation}
\textbf D=Y+f(Y,N;W_{unet}),
\end{equation}
where $W_{unet}$ denotes the network parameters of the denoising subnetwork. 

\begin{figure*}[htb]
\begin{center}
\includegraphics [width=7in]{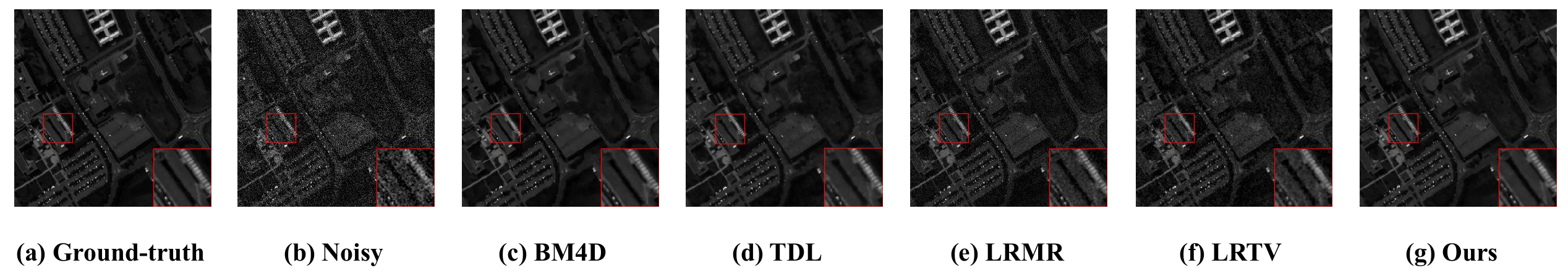}
\caption{Denoising results at $10^{th}$ band of image under noise level $\sigma=30$ on Pavia University dataset. }
\end{center}
\end{figure*}

\subsection{Loss Function}

In this paper, feature extracting and denoising are in an end-to-end framework. Thus, mean squared error (MSE) loss and perceptual loss are employed as the basic loss function to guide the learning of the denoising model.

The MSE loss is formulated as follows:
\begin{equation}
\mathcal {L}_{M}=MSELoss(G,D)={\textstyle\frac1{CHW}}\sum_{t=1}^{CHW}{(G_t-D_t)}^2,
\end{equation}
where $G$ and $D$ denote the ground truth and the denoised image respectively, $CHW$ denotes the number of pixels, $G_t$ and $D_t$ denote the ground truth at pixel $t$ and the denoised image at pixel $t$ respectively.

Furthermore, the perceptual loss is formulated as follows:
\begin{equation}
\mathcal {L}_{P}=\ell^{\phi,j}(G,D)=\frac1{C_jH_jW_j}{\parallel\phi_j(G)-\phi_j(D)\parallel}_2^2,
\end{equation}
where $\phi_j(\cdot)$ denote the $j$th layer of VGG-19. Therefore, the perceptual loss obtains more detailed information through the feature reconstruction of CNN.

In addition, we exploit asymmetric loss as a noise estimation loss, which is introduced from CBDNet \cite{Guo_2019_CVPR} to measure the noise estimation. Over-denoising or under-denoising are penalized by the asymmetric loss. The asymmetric loss is formulated as follows:
\begin{equation}
\mathcal {L}_{asymm}=\sum_{i}| \alpha - \mathbb{I}_{( N_{i}-N_{i}^{'} )<0} | \cdot(N_i-N_i^{'})^2,
\end{equation}
where ${\mathbb{I}}_e$ denotes the indicator function, $N_{i}$ and $N_{i}^{'}$ denote the noise level estimation at pixel $i$ and the ground truth at pixel $i$, respectively. $\alpha$ is empirically set to 0.25.

To sum up, the total loss function is given by:
\begin{equation}
\mathcal {L}={\mathcal L}_M+{\mathcal L}_P+\lambda_{asymm}{\mathcal L}_{asymm},
\end{equation}
where $\lambda_{asymm}$ denotes the parameters for the asymmetric loss.

\section{Experiments and Analysis}

\subsection{Experimental Setup}

We conduct the training process on the ICVL dataset, which is comprised of 201 images. The images in ICVL dataset were collected at $1392\times1300$ spatial resolution over 31 spectral channels. In order to expand the training dataset, each training image was cropped into multiple patches of size $64\times64\times31$. Furthermore, we use Pavia Center dataset to fine-tune the model. As for the testing part, we evaluate our model in Pavia University dataset. These two datasets were acquired by the ROSIS sensor and after processing, the size of Pavia Center image is $1096\times715\times102$ while the size of Pavia University image is $610\times340\times103$.

The Adam optimizer with batch size of 64 is employed to optimize the proposed method. The learning rate was initialized to $10^{-4}$ and we use a weight decay of 0.0005. The network was trained with 100 epochs based on the above settings. As for noise settings, the adding noise was referred as two cases: 1) adding AWGN with noise levels of 30, 50, 70; 2) randomly adding AWGN with the noise level ranging from 10 to 70.

\begin{table*}[htb]
\centering
\caption{Quantitative assessment results of different methods under several noise levels on Pavia University dataset. ``Blind" means corrupted by Gaussian noise with unknown $\sigma$ at each band.}
\begin{tabular}{c c|c c c c c c} \hline
Noise & Index & ~~Noisy~~ & ~~BM4D~~ & ~~TDL~~ &  ~~LRMR~~ &  ~~LRTV~~  & Proposed \\ \hline\hline
\multirow{3}*{$\sigma$ = 30} 
 & PSNR & 18.59 & 33.43 & 33.54 &27.14 &29.29 & \textbf{34.93} \\

~ & SSIM &0.193 &0.854 &0.849 & 0.578 &0.694 &\textbf{0.915}\\
~ & SAM  &0.875 &0.173 &0.152 &0.407 &0.326 & \textbf{0.102}\\ \hline
\multirow{3}*{$\sigma$ = 50} 
&PSNR&14.16 & 31.70 &31.81 &24.01 &26.25 &\textbf{32.46}\\

~ & SSIM &0.083 & 0.784 &0.788 &0.436 &0.531 &\textbf{0.874}\\
		
~ & SAM  &1.093 &0.207 &0.146 &0.613 &0.498 &\textbf{0.127}\\ \hline 
\multirow{3}*{$\sigma$ = 70} 
&PSNR& 11.23 &30.12 &30.38 &22.07 &24.18 &\textbf{30.68} \\

~ & SSIM & 0.045 &0.731 & \textbf{0.751} &0.232 &0.298 &0.743\\

~ & SAM  &1.207 &0.238 &0.186 &0.695 &0.545 &\textbf{0.154}\\ \hline 
\multirow{3}*{blind}
&PSNR& 17.14 &31.16 &26.86 &24.63 &28.86 &\textbf{32.81}\\

~ & SSIM &0.186 & 0.734 & 0.532 &0.469 &0.595 &\textbf{0.869}\\

~ & SAM  &1.048 &0.330 &0.496 &0.777 &0.454 &\textbf{0.124 }\\ 		
\hline
\end{tabular}
\label{gauss case}
\end{table*}

\subsection{Results and Analysis}

We compare the proposed method with other four closed related denoising methods: BM4D \cite{6253256}, TDL \cite{Peng_2014_CVPR}, LRMR \cite{6648433} and LRTV \cite{7167714} on Pavia University dataset. In order to evaluate the performance of these denoising methods, we use three quantitative evaluation indexes: PSNR, SSIM and SAM. Generally speaking, the higher values of PSNR and SSIM mean the better denoising effect, while the lower value of SAM means the better denoising effect.

Table 1 lists the quantitative assessment results of our method and other closely related denoising methods in different noise levels. As shown in the table, our method achieves better performance in quantitative evaluation. Fig.3 shows the visual comparison between our method and other denoising methods. The comparison showed that our method can effectively remove the noise while preserving the details of the image.

\section{ CONCLUSION}
\label{sec:typestyle}

In this paper, we propose a novel blind denoising method for HSIs based on multistream denoising network (MSDNet), which consists of noise estimation subnetwork and denoising subnetwork. The MSDNet can estimate the noise level autonomously, and then realize blind denoising and improve the performance of HSI denoising. The comparison with other methods indicates that the proposed method achieves better denoising performance under different noise levels.

In the future, we are committed to investigating hyperspectral images denoising under mixed noise, such as impulse noise and stripe noise. On the other hand, we will work on improving the accuracy of noise level estimation to improve the denoising process.

\end{document}